# Undulated 2D materials as a platform for large Rashba spin-splitting and persistent spin-helix states

Sunny Gupta[1], Manoj N. Mattur[1], and Boris I. Yakobson[1,2,3*]

[1]Department of Materials Science and Nanoengineering, Rice University, Houston, TX, 77005 USA
[2]Department of Chemistry, Rice University, Houston, TX 77005, USA
[3]Smalley-Curl Institute for Nanoscale Science and Technology, Rice University, Houston, TX, 77005 USA

## Abstract

Materials with large unidirectional Rashba spin-orbit coupling (SOC), resulting in persistent-spin helix states with small spin-precession length, are critical for advancing spintronics. We demonstrate a design-principle achieving it through specific undulations of 2D materials. Analytical model and first-principles calculations reveal that bending-induced asymmetric hybridization brings about and even enhances Rashba SOC. Its strength $α_R ∝ κ$ (curvature) and shifting electronic levels $Δ ∝ κ^2$. Despite vanishing integral curvature of typical topographies, implying net-zero Rashba effect, our two-band analysis and electronic structure calculation of a bent 2D MoTe$_2$ show that only an interplay of $α_R$ and $Δ$ modulations results in large unidirectional Rashba SOC with well-isolated states. Their high spin-splitting ~0.16 eV, and attractively small spin-precession length ~1 nm, belong among the best known. Our work uncovers major physical effects of undulations on Rashba SOC in 2D materials, opening new avenues for using their topographical deformation for spintronics and quantum computing.

Finding materials exhibiting large Rashba spin-orbit coupling (SOC) [1] – a relativistic effect causing spin-split electronic bands even in non-magnetic systems – is pivotal for advancing spintronics [2,3]. Rashba SOC is fundamental to many spin-based logic and memory devices [2,4], allowing spin-currents' electrical control in spin field-effect [5,6] and spin Hall [7,8] transistors, magnetization directions' electrical control via spin-orbit torque [9], and also quantum computations with Majorana fermions [10]. Rashba SOC exhibited by material with broken space inversion symmetry is generally described by the model Hamiltonian $H_R = α_R(\mathbf{k} \times \hat{\mathbf{z}}) \cdot \boldsymbol{\sigma} = \boldsymbol{\Omega}_R(\mathbf{k}) \cdot \boldsymbol{\sigma}$, [11] where $\boldsymbol{\sigma}$ is the Pauli matrix vector, parameter $α_R$ (incorporating all material specifics) represents the Rashba effect strength, $\hat{\mathbf{z}}$ indicates inversion asymmetry direction, and $\boldsymbol{\Omega}_R(\mathbf{k})$ is the momentum-dependent effective magnetic field which splits electronic bands into spin-polarized states with spin-momentum locking. An electron moving in such materials exhibits spin precession of length $L_{pr} \propto 1/α_R$, enabling spin-current modulation without external magnetic field. While $\boldsymbol{\Omega}_R(\mathbf{k})$ allows manipulating electron spin, its dependence on both magnitude and direction of $\mathbf{k}$ causes spin-dephasing by momentum scattering and limits device performance. Therefore, materials with large $α_R$ and unidirectional momentum-independent field $\boldsymbol{\Omega}_R(k)$ are needed to realize persistent-spin helix (PSH) [12,13] states with small $L_{pr}$ and large spin-diffusion length, crucial for achieving spin transport in the diffusive regime. While Rashba SOC has been predicted and observed in several materials [14–22], the $\boldsymbol{\Omega}_R(\mathbf{k})$ was not unidirectional, and $L_{pr}$ ~1 µm in devices was large, due to small $α_R$, limiting operating temperature [6]. Unidirectional $\boldsymbol{\Omega}_R(\mathbf{k})$ requires stringent symmetry [13], and only recently, PSH states with smaller $L_{pr}$ ~ 10 nm were predicted in a few bulk materials [23,24]; though without transport measurements. PSH states with small $L_{pr}$ are yet to be observed in two-dimensional (2D) materials, particularly promising [2,3] for miniaturization and complex spintronic device architectures using 2D heterostructures. Discovering 2D or bulk materials with the desired $\boldsymbol{\Omega}_R(\mathbf{k})$ and large $α_R$ is often serendipitous with complex synthesis procedures, making a physical (not chemical-synthetic) design strategy imperative.

Here, we demonstrate a design-principle to create large and unidirectional Rashba SOC in 2D materials by its undulation, bending with zero Gaussian curvature, avoiding in-plane strain, which otherwise lacks Rashba spin-splitting. Bending graphene [25] affects SOC, however it is weak due to carbon's low atomic number. More generally, deforming a 2D material in a topographical shape $z = f(x)$ with local bending curvature $κ(x) = f''/(1 + f'^2)^{3/2}$ (Fig. 1a) breaks inversion symmetry, introducing a potential gradient $\nabla V$ and effective pseudo-electric field $\mathbf{E}$ along the local normal direction. Electrons experiencing SOC in such a local $\mathbf{E}$ will further display Rashba spin-splitting, where $\boldsymbol{\Omega}_R(k)$ direction depends on the local $\mathbf{E}$. Thus, surface topography determines the SOC field $f(x) \rightarrow \boldsymbol{\Omega}_R(\mathbf{k})$, potentially enabling a unidirectional $\boldsymbol{\Omega}_R(\mathbf{k})$ through topographical undulations. Using analytical modeling and first-principles calculations, we uncover the physical effects of undulation on Rashba SOC in 2D materials; while creating Rashba SOC with strength $α_R \propto κ$, undulation also shifts electronic levels $\Delta \propto κ^2$. It may seem that for typical topographies (Fig. 1a) $\int κ(x)dx = 0$, and therefore net $α_R \propto κ$ will vanish. However, using a two-band model (2BM) and examining the electronic states of a bent 2D material, we show that the complex interplay of $α_R$ and $\Delta$ dependencies on local $κ$ determines the non-linear mapping between $f(x)$ and $\boldsymbol{\Omega}_R(\mathbf{k})$ and leads to non-vanishing Rashba effect with isolated large spin-split states, vital for efficient spin-selective transport. Our calculated electronic structure of 2D MoTe$_2$ in a feasible Gaussian-shaped deformation demonstrate isolated Rashba states with large unidirectional $\mathbf{\Omega_R}$, a small $L_{pr}$ ~1 nm, and high spin-splitting of ~0.16 eV (~6 $k_BT$ at room temperature), among the best for any known bulk or 2D materials. Our approach, leveraging the high bendability of 2D materials [26,27] due to their large Föppl–von Kármán number [26], opens a new area of utilizing topographical deformation for designing large and unidirectional Rashba SOC, with promise in spintronics [2,3] and quantum computing [10].

*Effects of undulations on electronic states in 2D materials* - The linear Rashba SOC in 2D systems is described by $H_R = \alpha_R(\mathbf{k} \times \hat{\mathbf{z}})\cdot\mathbf{\sigma}$, which splits 2D parabolic bands into spin-polarized states (Fig. 1b) given by $E(k) = \hbar^2 k^2/2m^* \pm \alpha_R k$ with spin-momentum locking – spin direction perpendicular to the wave vector. The Rashba parameter $\alpha_R$, representing effects' strength, is given by [28]

$$\alpha_R = \int d^3r \frac{1}{c^2} \frac{\partial V(r)}{\partial z} |\psi(r)|^2 \tag{1}$$

where $V(r)$ and $\psi(r)$ are the single-body potential and wave function, respectively. The asymmetry in $V$ and/or $|\psi|^2$ results in Rashba SOC. As a relativistic effect, SOC is dominant only close to the nuclei, where the antisymmetric Coulomb term $\nabla_z V$ is large in Eq. (1), so that Rashba effect arises primarily from the wavefunction's asymmetry if the inversion-symmetry is broken. [28,29] Bending even a symmetric 2D material into a shape $z = f(x)$, breaks its local inversion symmetry, by creating asymmetric hybridization. Generally, the wavefunction in the flat configuration localized on atomic orbital $\psi_{flat} = \psi_0$ will get modified to $\psi_{bent} = \psi_0 + const\cdot\psi_b \kappa$ due to orbital mixing from bending, for small bending angles [30]. Thus, from Eq. 1, in bent 2D materials with induced asymmetric hybridization (more details in Supplementary Information SI-1):

$$\alpha_R \propto \kappa \tag{2}$$

Concurrently, bending a 2D material creates a local strain $\varepsilon = z\kappa$, where $z$ is the distance along the bending direction within the layer's thickness. Using perturbation theory, the effect of bending on electronic levels can be described by Hamiltonian $H = H_0 + H_{strain} = H_0 + Dz\kappa$, with $H_0$ as the unperturbed state, and $D$ is a strain-induced deformation potential. Since $H_{strain}$ is an odd function in $z$, for small $\kappa$, the energy level shift $\Delta$ due to bending is (more details in SI-2):

$$\Delta \propto \kappa^2 \tag{3}$$

Thus, bending a 2D material should induce Rashba effect of magnitude $\alpha_R \propto \kappa$, and importantly a concurrent energy level shift $\Delta \propto \kappa^2$, the later as discussed subsequently is critical to get isolated Rashba states and non-zero Rashba effect for typical topographies.

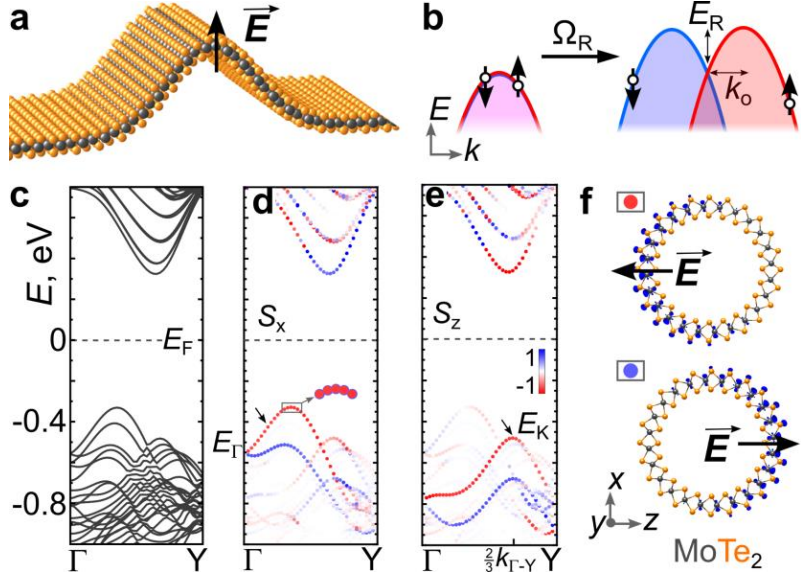

Fig 1. (a) Bent 2D material experiencing local pseudo-electric field $\vec{E}$ due to curvature-induced inversion symmetry breaking. (b) Schematic showing spin (up/down arrows) degenerate parabolic bands split into spin-polarized states due to Rashba SOC field $\Omega_R$. $E_R$ and $k_o$ denote Rasha energy and momentum-splitting, respectively. (c) Electronic band structure of (12,12) MoTe$_2$ nanotube, with projected values of $S_x$ in (d) and $S_z$ in (e), where inset shows a color map. The arrow in (d) points to Rashba spin-split band with eigenvalue $E_\Gamma$ at Γ point, while in (e) points to VBM bands $E_K$ at K-point in flat 2D MoTe$_2$, which folds at ⅔ Γ-Y in armchair nanotubes. (f) Band-decomposed charge density (blue) of spin-degenerate VB states, red and blue in gray box in (d). The effective $\vec{E}$ direction at the localized states is also shown.

*Dependence of Rashba SOC and Δ on κ* - To determine the scale and ascertain the behavior of $\alpha_R$ and Δ with curvature (which always refers to bending) in different 2D materials, we used density functional theory (DFT) with SOC (details in Methods) to calculate the electronic structure in a cylindrical nanotube (NT) geometry of different radii $R$, with uniform $\kappa = 1/R$. The 2H phase [31] MoSe$_2$, MoTe$_2$, and WTe$_2$ with high atomic number $Z$, are experimentally common and chosen as prototypes, in armchair NT configuration (periodic along $y$) with chiral indices ($n$, $n$). Fig. 1c shows the electronic bands of (12,12) MoTe$_2$ NT along Γ−Y, with SOC (with no SOC see Fig. S2a), indicating a semiconductor. To reveal Rashba spin-splitting, we plot the electronic structure with projected expectation values of spin operators $<S_x>$ and $<S_z>$ in Fig. 1d-e ($<S_y>$ are negligible, shown in Fig. S2c). The red and blue bands representing opposite spin values show spin-splitting. The valence band maxima (VBM) states have major contributions from $<S_x>$ and are spin-degenerate (there are blue dots behind red dots). We also plot in Fig. 1f the spatial wave function localization $|\psi|^2$ of the degenerate VB states shown in gray box in Fig. 1c. The states corresponding to red-bands (blue-bands) are localized on the left (right) of the tube due to circular confinement. For the VBM states, spatial localization and local structural asymmetry along the $z$-direction mean that from $H_R = \alpha_R(\boldsymbol{k_y} \times \hat{\boldsymbol{z}})\cdot\boldsymbol{\sigma}$, only $<S_x>$ will be dominant. This is evident in Fig. 1d indicating Rashba spin-splitting in the VBM states.

We further extract the Rashba parameter $\alpha_R$ from the electronic band structure. It is routinely estimated using the formula $\alpha_{R,p} = 2E_R/k_o$ ($E_R$ and $k_o$ are shown in Fig. 1b), applicable only for parabolic, free electron-like bands and linear Rashba SOC. However, for the transition metal dichalcogenide (TMD) of $D_{3h}$ symmetry, one should use the model Hamiltonian (mH) [32] with higher-order terms in $k$,

$$H_R(k_y) = \alpha_{R,mH} k_y \sigma_x + \alpha_{3R} k_y^3 \sigma_x + \alpha_{3R'} k_y^3 \sigma_z, \qquad (4)$$

where $\alpha_{3R}$ and $\alpha_{3R'}$ are cubic coupling constants. Fitting the Rashba spin-split bands (Fig. 1d) with this model Hamitonian yields $\alpha_{R,mH}$ = 0.41 eVÅ, $\alpha_{3R}$ = −1.47 eVÅ$^3$, and $\alpha_{3R'}$ = −1.53 eVÅ$^3$ (fitting details in SI-3). While the routine formula yields $\alpha_{R,p} = 2E_R/k_o$ = 1.45 eVÅ, from $E_R$ = 225 meV and $k_o$ = 0.31/Å, higher than the accurate $\alpha_{R,mH}$ due to $k^3$ terms in the mH, and mexican hat-like dispersion of electronic bands without SOC (Fig. S2), which overestimates $E_R$. We similarly extract $\alpha_R$ for different 2D TMDs in NT geometries, and Fig. 2a shows $\alpha_{R,mH}$ and $\alpha_{R,p}$ as a function of $\kappa$, well fitting linear $\alpha_R \propto \kappa$ at small $\kappa$, as expected from Eq. 2. At higher $\kappa$, the $k^3$ cubic coupling terms become large (Fig. S3), which possibly makes $\alpha_{R,mH}$ depend non-linearly on $\kappa$. The slope of $\alpha_{R,mH}$ vs $\kappa$ is highest for WTe$_2$ (Fig. 2a) due to the higher atomic numbers $Z$ of W ($Z$ = 74) and Te ($Z$ = 52). It should be noted that $\alpha_{R,p}$ is also proportional to $\kappa$, however the fitted slopes for MoTe$_2$ and WTe$_2$ are similar. This is not expected as SOC strength should increase with $Z$, indicating that routinely used $\alpha_{R,p}$ doesn't capture essential physics. An important metric for device design is spin-precession length $L_{pr} = \pi/\delta k$, where $\delta k$ is the wavenumber difference between two Rasbha bands. TMD nanotubes with radius of $R$ = 1.5 nm have been experimentally synthesized [27], and for such WTe$_2$ tubes with $\kappa$ = 0.07/Å, $L_{pr}$ = 1.2 nm, which is among the lowest spin-precession length known among all bulk/2D materials [24] showing Rashba SOC.

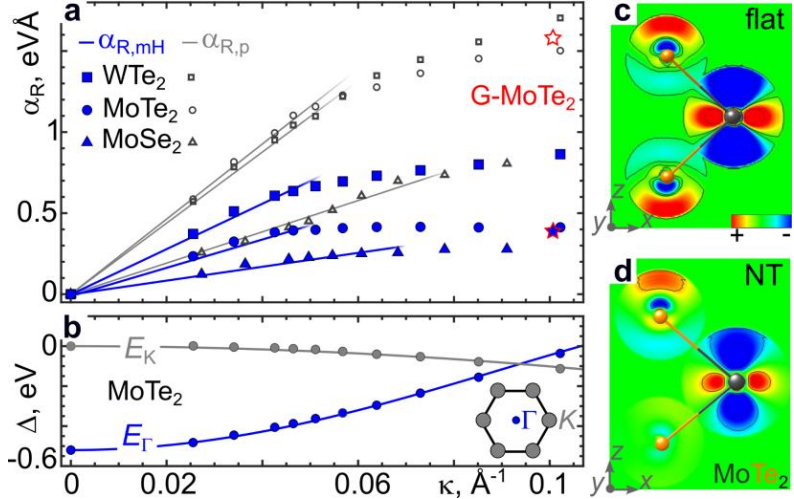

Fig 2. (a) Rashba parameter $\alpha_{R,mH}$ (blue line, filled symbols) and $\alpha_{R,p}$ (gray line, empty symbols) as a function of nanotube curvature for different TMDs. G-MoTe$_2$ (star) is the Gaussian curve in Fig. 3a. Solid lines show linear fit $\alpha_{R,mH}, \alpha_{R,p} \propto \kappa$. (b) Energy level shift $\Delta$ of eigenvalue $E_K$ (gray) and $E_\Gamma$ (blue) in MoTe$_2$ tubes of different radii. Solid lines fit $\Delta \propto \kappa^2 + \kappa^4$. Inset shows constant energy contour of valence bands (hole-valleys near $K$ and $\Gamma$) in flat 2D MoTe$_2$. Planar cross-section of the all-electron spinor wave function of Rashba states in the core-region for flat 2D MoTe$_2$ (c) and for (12,12) MoTe$_2$ nanotube (d).

Apart from large Rashba spin-splitting, for better spin-selective transport and device performance these bands must be isolated near the valence and/or conduction band edges without other degenerate bands. Fortuitously, the Rashba bands near Γ-point in (12,12) MoTe$_2$ NT (Fig. 1d) are isolated, however the Γ valley states are not the VBM in flat 2D TMDs (Fig. S4). The VB in flat 2D TMDs consists of $K$ and Γ valley (inset Fig. 2b), where the VBM is at $K$ point, which folds at $k = ⅔$ Γ-Y in armchair NT (Fig. 1e), termed $E_K$. While the Rashba bands occur at Γ-point in TMD NTs, termed $E_Γ$ (Fig. 1d). In flat 2D TMDs, $E_Γ$ is 0.52 eV below $E_K$ (Fig. S4), and bending induces band shifts (Eq. 3) and pushes Rashba states in (12,12) MoTe$_2$ NT to VBM. To illustrate the shift in energy levels due to bending, we show the evolution of eigenvalue $E_K$ and $E_Γ$ in MoTe$_2$ NTs of different radii in Fig. 2b, with $E_K$ in the flat layer set as zero. The solid lines (Fig. 2b) represent the expected analytical fit $Δ ∝ κ^2 + κ^4$ (Eq. 3). At small $κ$, the Rashba states are not at VBM, appearing there for tubes with $R < 1.5$ nm, as also seen in (12,12) MoTe$_2$ (Fig. 1d) – an effect of bending induced energy level shift (see Fig. S4 for the evolution of Rashba bands with $κ$ in MoTe$_2$ NTs). Rashba SOC caused by undulations, as described earlier (Eq. 1), arises due to bending-induced asymmetric hybridization. To demonstrate this, we show in Fig. 2d a planar cross-section of the all-electron spinor wave function in the core-region (near nuclei, where SOC is active) for Rashba states (gray box in Fig. 1d) in (12,12) MoTe$_2$ NT. The plane considered is perpendicular to the tube, passing through Te-Mo-Te atoms. For comparison, a similar plot is shown in Fig. 2c for flat 2D MoTe$_2$. The wavefunction is localized on all the atoms in the flat layer, while if curved it is localized only on Mo and the outer Te atoms, and expressly near their heavy nuclei, signifying asymmetric hybridization. This reveals that bending induced asymmetry in the wavefunction is responsible for Rashba SOC.

Although certain TMD NTs show isolated Rashba bands with large spin-splitting, the Rashba spin-split VBM states in Fig. 1d are spin degenerate with opposite values of $<S_x>$ due to the axial symmetry of the NTs $∫κ(φ)dφ = 0$. These states are spatially localized in the z-direction along the cylindrical arc with opposite local $κ$. From $H_R = α_R(\mathbf{k_y} × \hat{\mathbf{z}})·\mathbf{σ}$, the effective $\mathbf{Ω}(k)$ field points in opposite directions, $+x$ ($-x$) for left (right) states with degenerate energy. This spin degeneracy is detrimental for spin-current modulation in devices due to spin-dephasing. It may seem that in a 2D material bent into practical topographies, "flat on both left and right" (Fig. 1a), the integral $∫κ(x)dx = 0$, suppressing overall Rashba spin-splitting. However, unlike NTs, 2D materials can be bent into shapes with non-uniform $κ$. As will be shown below, in such shapes $z = f(x)$, although $∫κ(x)dx = 0$, the net effect of spin-splitting due to bending will be non-zero due to the interplay between Rashba SOC and $Δ$.

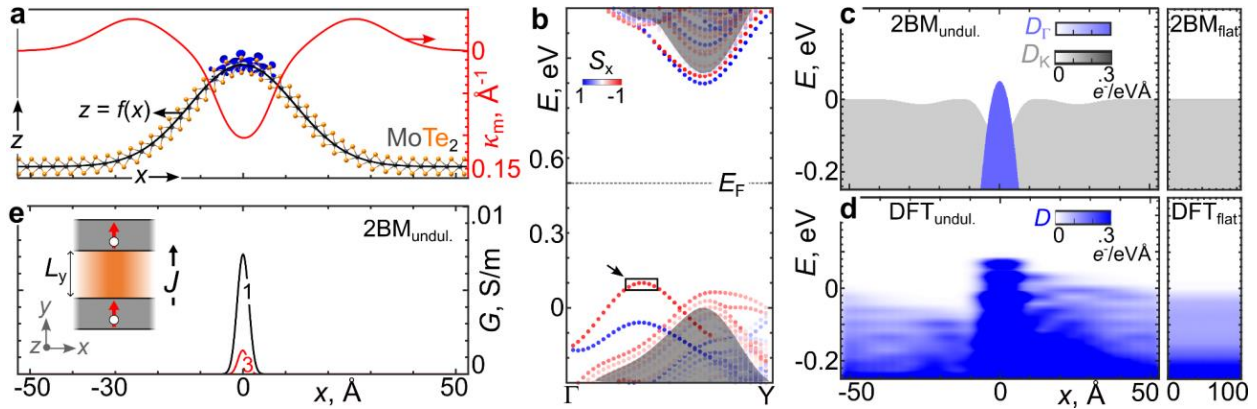

Fig 3. **(a)** Relaxed atomistic structure of G-MoTe$_2$ with band-decomposed charge density (blue) of Rashba states near VBM, arrow in (b). Black line follows $z = A·\exp(-x^2/2B^2)$, where $A = 24.03$ Å, $B = 12.4$ Å. Red line shows the local mean-curvature $κ_m$ of the "atomized" Gaussian. **(b)** Electronic band structure of G-MoTe$_2$ with projected $S_x$ values. Arrow points to Rashba bands. Gray shaded regions mark the bands of flat MoTe$_2$. **(c)** Evolution of local density of states $D$ of $K$ (Γ)-valley states in gray

(violet) calculated using a two-band model (2BM) in undulated G-MoTe$_2$ (left panel) and flat MoTe$_2$ (right panel). **(d)** Similar plot as (c) calculated using DFT. **(e)** Conductance in G-MoTe$_2$ calculated using 2BM for $L_y$ = 1 or 3 nm (black or red). Inset shows device setup with ferromagnetic-electrodes (gray) and transport channel (orange). Red arrow is the electron-spin direction. Current $J$ flows along $y$.

*2-band model (2BM) of electronic states in an undulated 2D material* - Having determined the dependence of α$_R$ and Δ on curvature $\kappa$, we resort to a 2BM to examine the electronic states of a 2D material bent in a generic shape $z = f(x)$. 2BM includes a hole-valley at $K$ point with parabolic dispersion $E_p(k) = -½\hbar^2k^2 m_K$ and a Rashba spin-split valley at Γ point $E_R(k) = -½\hbar^2k^2 m_\Gamma ± α_R k + E_0$, where $E_0$ is the relative energy difference between the two valleys in flat layer (Fig. S6a,b). This is representative of the valence bands in 2D flat TMDs, as shown by the constant-energy contour in Fig. 2b inset. To examine how the electronic states change by bending, we derive (see SI-4) the macroscopically-averaged local density-of-states (LDOS) $D(E,x)$ of our model in a shape of local curvature $\kappa(x)$. For a long-scale undulation, neglecting quantization effects, the LDOS is a sum of two contributions, $D_p$ and $D_R$

$$D_p(E,x) \cong \frac{L_y m_K}{\pi \hbar^2} \text{ for } E < \Delta_K(x) \tag{5}$$

$$D_R(E,x) \cong \frac{L_y m_\Gamma}{\pi \hbar^2} \left(\frac{E_R(x)}{E_{R,m}(x)-E}\right)^{1/2} \text{ for } \Delta_{\Gamma m}(x) \le E < E_{Rm}(x), \text{ or}$$

$$\cong \frac{L_y m_\Gamma}{\pi \hbar^2} \text{ for } E \le \Delta_{\Gamma m}(x) \tag{6}$$

where p or R label the parabolic or Rashba bands, $α_R(x) = c_{1,R}\kappa(x) + c_{2,R}\kappa(x)^2 + c_{3,R}\kappa(x)^3$, $\Delta_{Kor\Gamma}(x) = c_{1,s}\kappa(x)^2 + c_{2,s}\kappa(x)^4$, $E_R(x) = m_\Gamma α_R(x)^2/2\hbar^2$, $c_{i,R(s)}$ are material-specific constants, $\Delta_{\Gamma m}(x) = \Delta_\Gamma(x) + E_0$, $E_{Rm}(x) = E_R(x) + \Delta_{\Gamma m}(x)$, $L_y$ is a unit-cell size along $y$, and $D_p$, $D_R$ are zero outside the $E$ ranges. As one could expect, Eq. 6 includes van Hove singularity feature.

The effect is illustrated for a Gaussian shape $z(x) = A \cdot \exp(-½x^2/B^2)$, with $A$, and $B$ for amplitude, and width, Fig. 3a. Although the function is smooth, the lattice is atom-structured, so its effective local mean-curvature experienced by material is defined as $\kappa_m(x) = \int_{x-a/2}^{x+a/2} \kappa(x)dx/a$ ($a$ = 3.55 Å, lattice constant of flat MoTe$_2$). The red line in Fig. 3a depicts variations of $\kappa_m(x)$, largest $\kappa_{m,max}$ ~ 0.1/Å at the tip. Fig. 3c shows $D(E,x)$ for Γ (blue) or $K$ (gray) hole-valley states, calculated with Eqs. 5-6 for flat (right panel) as well for Gaussian shape (left panel), with energy referenced to the VBM position in the flat layer. Parameters for 2D MoTe$_2$ [33] and the shape function shown in Fig. 3a (black curve) were used. The values for $α_R(x)$ and $\Delta(x)$ as a function of $\kappa(x)$ were obtained [33] by analytically fitting their corresponding values for MoTe$_2$ from NT geometries (Fig. 2a,b). In the flat layer, $D(E,x)$ is of course uniform with $x$, while the VBM arises only from the $K$-valley. With an undulation, $D(E,x)$ varies depending on the local $\kappa(x)$. In the left and right flat regions $\kappa(x)$ is small, the valence band states still arise from the $K$-valley, while the Γ-valley Rashba states are deep below the VBM. However, in the middle region, due to high $\kappa$, the Rashba states shift upwards in energy, and spatially the VBM is located at the Gaussian tip with the highest $\kappa$. This 2BM analysis shows that the hole valleys modulation by layer curvature $\kappa(x)$ is twofold: curvature-induced Rashba spin-splitting $α_R(\kappa(x))$ and also local shift in energy levels $\Delta(\kappa(x))$ lifting the Rashba bands above the obscuring background, so they can appear distinctly at VBM for appropriately chosen shapes $z = f(x)$ with high $\kappa_{m,max}$.

To confirm the results of 2BM and the appearance of spin-split Rashba states at VBM in a factual 2D material, DFT with SOC were used to calculate the electronic structure of MoTe$_2$ deformed in the same Gaussian shape along the armchair direction, termed G-MoTe$_2$, Fig. 3a.

The atomistic structure in Fig. 3a is relaxed under shape-constraint (details in Methods). Its electronic band dispersion $E(k_y)$ with spinor projections $<S_x>$ is shown in Fig. 3b (similar to Fig. 1d, but notably with no "blue-red" degeneracy); other spinor components appear negligible for states near VBM (Fig. S5). Spatially, the VBM states (black box in Fig. 3b) are localized at the Gaussian tip (where its surface normal, the asymmetry direction, is along $\hat{z}$), as shown by $|\psi|^2$ plotted in Fig. 3a. From $H_R = \alpha_R(\mathbf{k}_y \times \hat{z})\cdot\mathbf{\sigma}$, only $<S_x>$ is non-zero for the VBM states, as also seen in Fig. 3b, that is the topmost valence bands are indeed Rashba spin-split states. From the band structure, we extract $\alpha_{R,mH}$ = 0.39 eVÅ, and $\alpha_{R,p}$ = 1.58 eVÅ (stars in Fig. 2a), consistent with that expected from their analytical values $\alpha_R(\kappa)$ [33], for $\kappa_{m,max}$ ~ 0.1/Å at the tip. In Fig. 3d, the $D(E,x)$ is computed with DFT for flat (uniform, right panel) and for curved G-MoTe$_2$ (expressly varying with $x$, left panel). The VBM is located near the Gaussian tip, and a comparison with Fig. 3b makes it apparent that those are Rashba states, similar to predictions of the 2BM. The spatial variation of $D(E,x)$, from DFT (Fig. 3d) and the 2BM (Fig. 3c) qualitatively agree. The electronic states follow the local $\kappa$-induced Rashba spin-splitting $\alpha_R(\kappa)$ and $\Delta(\kappa)$, the latter raising the Rashba states to VBM in G-MoTe$_2$. The lack of exact mirror symmetry in the armchair direction results in asymmetric $D(E,x)$ of G-MoTe$_2$ obtained by DFT (Fig. 3d), with additional horizontal-stripe-like features due to quantization. Nonetheless, both 2BM and DFT results indicate that the electronic states in a 2D material undulated by $z = f(x)$ can be understood and quantified by its associated $\kappa(x)$ and material-specific functions of $\alpha_R(\kappa)$ and $\Delta(\kappa)$.

Unlike in NTs, the Rashba spin-split states in G-MoTe$_2$ (Fig. 3b) are non-degenerate, with magnetic field $\mathbf{\Omega}_R(k) = \alpha_R(\mathbf{k}_y \times \hat{z})$ unidirectionally pointing along $+x$ direction, evinced by non-zero $<S_x>$ in Fig. 3b. Additionally, unlike in NTs, the non-uniform $\kappa$ of Gaussian shape shifts Rashba states to VBM only locally, at the Gaussian tip, where $\mathbf{E}$ field is along $z$, directing $\mathbf{\Omega}_R$ field along $x$; thus non-uniform bending is crucial for unidirectional $\mathbf{\Omega}_R$. Such unidirectional $\mathbf{\Omega}_R$ prevents spin-dephasing and is central in modulating current $J$ by spin-precession – backbone behind spin-FET devices [5,34]. We examine current modulation due to Rashba spin-splitting in a 2D layer undulated along $x$, in a device geometry of Fig. 3e inset. It contains ferromagnetic-electrodes (gray regions) with spin polarized along $y$, and a transport channel (orange region) of undulated 2D layer. In the ballistic limit, under small-bias voltage $V$, at temperature $T$, and neglecting quantization in $x$, the integral current (more details in SI-5) is

$$J = \int J(x)dx \cong 4VG_0 Ce^{-\beta F(0)}\cdot(2\pi/\beta F'(0))^{1/2}\cdot\cos^2\varphi \qquad (7)$$

where, $G_0 = 2e^2/h$, $C = m_\Gamma^*\alpha_{R,x=0}/\pi\hbar^2$, $\varphi = L_y m_\Gamma^*\alpha_{R,x=0}/\hbar^2$, $F(x) = E_F - E_{Rm}(x)$, $\beta = 1/k_B T$. Using parameters for G-MoTe$_2$ and $E_F$ = 0.5 eV (Fig. 3b), the conductance $G = J(x)/V$ is plotted in Fig. 3e, showing $x$-localization at the Gaussian tip due to bending-induced $\Delta$, which shifts the Rashba states to the VBM. Spin of electron entering the channel while pointing along $y$ will precess in field $\mathbf{\Omega}_R$, and so its cumulative exit-phase shift will define current $J$, which thus depends sensitively on channel length, $L_y$. Having known that an external $\mathbf{E}$ field can change $\alpha_R$, it will significantly modulate cumulative phase and the device current, as in spin-FET [5,6,19]. Similar effects are certainly expected in other deformations $z = f(x)$ with non-uniform $\kappa(x)$.

In conclusion, bending 2D TMDs results in large Rashba spin-splitting, highlighted by the concurrent band shift, and their complex interplay with curvature $\kappa$ ensures Rashba bands are well isolated near VBM, which is critical for efficient spin-selective transport in spin-FETs. Additionally, the intricate interplay of $\alpha_R$ and $\Delta$ with $\kappa$ in 1D Gaussian deformed MoTe$_2$ creates a unidirectional $\mathbf{\Omega}_R$ generating persistent spin-helix states [13,34], with no complex spin texture, which are pivotal for spin-transport in the diffusive regime and hitherto has been challenging to achieve. TMD NTs [27] with small radius ($\kappa \sim 0.07$/Å) and bent 2D materials [26] having high local curvature $\kappa \sim 0.1$/Å have been experimentally synthesized. Thus, achieving a local $\kappa_{max} = 0.1$/Å in 2D MoTe$_2$ is feasible. Such systems are expected to have a PSH state with small $L_{pr} \sim 1$ nm and spin-splitting reaching $E \sim 0.16$ eV, among the best known [24] for Rashba spin-splitting in any material. This work opens new possibilities of using topographical deformation in 2D materials for spintronics [2,3], spin-FETs [5,6], spin-orbit torques [9], spin valves [4,35], and quantum computing [36], adding exotic physical behaviors achievable in undulated 2D materials [37–41].

**Methods**

**Density functional theory (DFT) calculations:** First-principles density functional theory (DFT) calculations were performed using the Vienna Ab Initio Simulation Package (VASP) [42]. Ion-electron interactions were represented by all-electron projector augmented wave potentials [43]. The generalized gradient approximation (GGA) parameterized by Perdew-Burke-Ernzerhof [44] (PBE) was used to account for the electronic exchange and correlation. A plane wave basis with a kinetic energy cut-off of 500 eV was used for wave functions expansion. The structures were relaxed until the Hellmann-Feynman forces on the atoms were less than 0.01 eV/Å. A Monkhorst-pack grid of 1x21x1 *k*-points was used to sample the Brillouin zone in nanotubes and undulated 2D MoTe$_2$. A vacuum of 20 Å (30 Å) was used along the non-periodic direction in the nanotubes (undulated 2D MoTe$_2$) to reduce the interaction between the periodic images. Spin-orbit coupling (SOC) was included in all the calculations. The DFT-D3 method with Becke-Johnson damping function was applied to simulate the van der Waals interaction. The all-electron wavefunction in the PAW formalism was calculated using the VaspBandUnfolding package [45].

**Creating a Gaussian deformed structure:** To create the structure of 2D MoTe$_2$ deformed in a 1D Gaussian shape $z(x) = A \cdot \exp(-x^2/2B^2)$ (black curve in Fig. 3a), where *A*, and *B*, are amplitude, and width, we use a constrained relaxation method. Firstly, a 20x1 supercell of 2D MoTe$_2$ in a rectangular unit cell with *x*-axis along the armchair direction was created. To transform the flat layer into the 1D Gaussian structure the *x* coordinate in the flat layer was transformed to *x'*, such that *x'* lies on the 1D Gaussian surface. *x'* was calculated by numerically evaluating the perimeter of 1D Gaussian such that $x = \int_0^{x'} (1 + (\frac{\partial z(x)}{\partial x})^2)^{1/2} dx$. The *z*-coordinate was then transformed accordingly depending on the obtained value of *x'* by $z(x') = A \cdot \exp(-x'^2/2B^2)$. This was done for the top and bottom Te layers as well as the Mo layer. Subsequently, constrained relaxation using DFT was used to obtain the relaxed structure. Firstly, Mo atoms were fixed and Te atoms were fully relaxed, and vice versa. The structures were relaxed until the Hellmann-Feynman forces on the atoms were less than 0.01 eV/Å.

**References**

[1] E. Rashba, Properties of semiconductors with an extremum loop. 1. Cyclotron and combinational resonance in a magnetic field perpendicular to the plane of the loop., Sov Phys Solid State **2**, 1109 (1960).

[2] H. C. Koo et al., Rashba Effect in Functional Spintronic Devices, Adv. Mater. **32**, 2002117 (2020).

[3] A. Hirohata, K. Yamada, Y. Nakatani, I.-L. Prejbeanu, B. Diény, P. Pirro, and B. Hillebrands, Review on spintronics: Principles and device applications, J. Magn. Magn. Mater. **509**, 166711 (2020).

[4] A. Manchon, H. C. Koo, J. Nitta, S. M. Frolov, and R. A. Duine, New perspectives for Rashba spin–orbit coupling, Nat. Mater. **14**, 871 (2015).

[5] S. Datta and B. Das, Electronic analog of the electro-optic modulator, Appl. Phys. Lett. **56**, 665 (1990).

[6] P. Chuang et al., All-electric all-semiconductor spin field-effect transistors, Nat. Nanotechnol. **10**, 35 (2015).

[7] J. Sinova, D. Culcer, Q. Niu, N. A. Sinitsyn, T. Jungwirth, and A. H. MacDonald, Universal Intrinsic Spin Hall Effect, Phys. Rev. Lett. **92**, 126603 (2004).

[8] W. Y. Choi, H. Kim, J. Chang, S. H. Han, H. C. Koo, and M. Johnson, Electrical detection of coherent spin precession using the ballistic intrinsic spin Hall effect, Nat. Nanotechnol. **10**, 666 (2015).

[9] P. Noël et al., Non-volatile electric control of spin–charge conversion in a SrTiO3 Rashba system, Nature **580**, 483 (2020).

[10] J. D. Sau, R. M. Lutchyn, S. Tewari, and S. Das Sarma, Generic New Platform for Topological Quantum Computation Using Semiconductor Heterostructures, Phys. Rev. Lett. **104**, 040502 (2010).

[11] Y. A. Bychkov and E. I. Rashba, Properties of a 2D electron gas with lifted spectral degeneracy., JETP Lett **39**, 78 (1984).

[12] B. A. Bernevig, J. Orenstein, and S.-C. Zhang, Exact SU(2) Symmetry and Persistent Spin Helix in a Spin-Orbit Coupled System, Phys. Rev. Lett. **97**, 236601 (2006).

[13] L. L. Tao and E. Y. Tsymbal, Persistent spin texture enforced by symmetry, Nat. Commun. **9**, 1 (2018).

[14] J. Luo, H. Munekata, F. F. Fang, and P. J. Stiles, Effects of inversion asymmetry on electron energy band structures in GaSb/InAs/GaSb quantum wells, Phys. Rev. B **41**, 7685 (1990).

[15] J. Nitta, T. Akazaki, H. Takayanagi, and T. Enoki, Gate Control of Spin-Orbit Interaction in an Inverted In0.53Ga0.47As/In0.52Al0.48As Heterostructure, Phys. Rev. Lett. **78**, 1335 (1997).

[16] S. LaShell, B. A. McDougall, and E. Jensen, Spin Splitting of an Au(111) Surface State Band Observed with Angle Resolved Photoelectron Spectroscopy, Phys. Rev. Lett. **77**, 3419 (1996).

[17] K. Ishizaka et al., Giant Rashba-type spin splitting in bulk BiTeI, Nat. Mater. **10**, 521 (2011).

[18] Z.-H. Zhu et al., Rashba Spin-Splitting Control at the Surface of the Topological Insulator Bi2Se3, Phys. Rev. Lett. **107**, 186405 (2011).

[19] S. Gupta and B. I. Yakobson, What Dictates Rashba Splitting in 2D van der Waals Heterobilayers, J. Am. Chem. Soc. **143**, 3503 (2021).

[20] S. Singh and A. H. Romero, Giant tunable Rashba spin splitting in a two-dimensional BiSb monolayer and in BiSb/AlN heterostructures, Phys. Rev. B **95**, 165444 (2017).

[21] T. Hu, F. Jia, G. Zhao, J. Wu, A. Stroppa, and W. Ren, Intrinsic and anisotropic Rashba spin splitting in Janus transition-metal dichalcogenide monolayers, Phys. Rev. B **97**, 235404 (2018).

[22] L. Yuan, Q. Liu, X. Zhang, J.-W. Luo, S.-S. Li, and A. Zunger, Uncovering and tailoring hidden Rashba spin–orbit splitting in centrosymmetric crystals, Nat. Commun. **10**, 906 (2019).

[23] H. Lee, J. Im, and H. Jin, Emergence of the giant out-of-plane Rashba effect and tunable

[23] nanoscale persistent spin helix in ferroelectric SnTe thin films, Appl. Phys. Lett. **116**, 022411 (2020).
[24] L. Zhang et al., Room-temperature electrically switchable spin–valley coupling in a van der Waals ferroelectric halide perovskite with persistent spin helix, Nat. Photonics **16**, 529 (2022).
[25] D. Huertas-Hernando, F. Guinea, and A. Brataas, Spin-orbit coupling in curved graphene, fullerenes, nanotubes, and nanotube caps, Phys. Rev. B **74**, 155426 (2006).
[26] E. Han, J. Yu, E. Annevelink, J. Son, D. A. Kang, K. Watanabe, T. Taniguchi, E. Ertekin, P. Y. Huang, and A. M. van der Zande, Ultrasoft slip-mediated bending in few-layer graphene, Nat. Mater. **19**, 305 (2020).
[27] Y. Nakanishi et al., Structural Diversity of Single-Walled Transition Metal Dichalcogenide Nanotubes Grown via Template Reaction, Adv. Mater. **35**, 2306631 (2023).
[28] M. Nagano, A. Kodama, T. Shishidou, and T. Oguchi, A first-principles study on the Rashba effect in surface systems, J. Phys. Condens. Matter **21**, 064239 (2009).
[29] G. Bihlmayer, P. Noël, D. V. Vyalikh, E. V. Chulkov, and A. Manchon, Rashba-like physics in condensed matter, Nat. Rev. Phys. **4**, 642 (2022).
[30] T. Dumitrică, C. M. Landis, and B. I. Yakobson, Curvature-induced polarization in carbon nanoshells, Chem. Phys. Lett. **360**, 182 (2002).
[31] J. Zhou et al., A library of atomically thin metal chalcogenides, Nature **556**, 355 (2018).
[32] Sz. Vajna, E. Simon, A. Szilva, K. Palotas, B. Ujfalussy, and L. Szunyogh, Higher-order contributions to the Rashba-Bychkov effect with application to the Bi/Ag(111) surface alloy, Phys. Rev. B **85**, 075404 (2012).
[33] $E_0 = -0.52$ eV, $m_K = 0.58 m_e$, $m_\Gamma = 6.61 m_e$, $\alpha_R(x) = 13.8\kappa(x) - 142.7\kappa(x)^2 + 456.3\kappa(x)^3$ eVÅ, $\Delta_\Gamma(x) = 59.8\kappa(x)^2 - 1233\kappa(x)^4$ eV, $\Delta_K(x) = -10.1\kappa(x)^2 + 0.08\kappa(x)^4$ eV, $E_F = 0.5$ eV, where $\kappa(x)$ is in 1/Å.
[34] J. Schliemann, J. C. Egues, and D. Loss, Nonballistic Spin-Field-Effect Transistor, Phys. Rev. Lett. **90**, 146801 (2003).
[35] Z. Kovács-Krausz et al., Electrically Controlled Spin Injection from Giant Rashba Spin–Orbit Conductor BiTeBr, Nano Lett. **20**, 4782 (2020).
[36] F. Wilczek, Majorana returns, Nat. Phys. **5**, 614 (2009).
[37] J. Mao et al., Evidence of flat bands and correlated states in buckled graphene superlattices, Nature **584**, 7820 (2020).
[38] R. Banerjee, V.-H. Nguyen, T. Granzier-Nakajima, L. Pabbi, A. Lherbier, A. R. Binion, J.-C. Charlier, M. Terrones, and E. W. Hudson, Strain Modulated Superlattices in Graphene, Nano Lett. **20**, 3113 (2020).
[39] S. P. Milovanović, M. Anđelković, L. Covaci, and F. M. Peeters, Band flattening in buckled monolayer graphene, Phys. Rev. B **102**, 245427 (2020).
[40] S. Gupta, H. Yu, and B. I. Yakobson, Designing 1D correlated-electron states by non-Euclidean topography of 2D monolayers, Nat. Commun. **13**, 3103 (2022).
[41] H. Yu, A. Kutana, and B. I. Yakobson, Electron Optics and Valley Hall Effect of Undulated Graphene, Nano Lett. **22**, 2934 (2022).
[42] G. Kresse and J. Furthmüller, Efficiency of ab-initio total energy calculations for metals and semiconductors using a plane-wave basis set, Comput. Mater. Sci. **6**, 15 (1996).
[43] P. E. Blöchl, Projector augmented-wave method, Phys. Rev. B **50**, 17953 (1994).
[44] J. P. Perdew, K. Burke, and M. Ernzerhof, Generalized Gradient Approximation Made Simple, Phys. Rev. Lett. **77**, 3865 (1996).
[45] https://github.com/QijingZheng/VaspBandUnfolding